\documentclass[11pt,preprintnumbers,a4paper]{article}
\usepackage{jheppub}
\usepackage{bbm}
\usepackage{epsfig,bm}
\usepackage{graphics,graphicx,comment,subfigure}
\usepackage{amsmath,booktabs}

\newcommand{\pslash}{\not{\hbox{\kern-2.3pt $p$}}}
\newcommand{\qslash}{\not{\hbox{\kern-2.3pt $q$}}}
\newcommand{\kslash}{\not{\hbox{\kern-2.3pt $k$}}}
\newcommand{\partialslash}{\not{\hbox{\kern-2.3pt $\partial$}}}

\def \be { \begin{equation} }
\def \ee { \end{equation} }

\def\tr{{\mbox{Tr}}}

\def\Dslash{{\rlap{\raise 1pt \hbox{$\>/$}}D}}

\def \( {\left(}
\def \) {\right)}

\def\slashchar#1{\ensuremath{                               %
   \setbox0=\hbox{${}#1{}$}       
   \dimen0=\wd0                                 
   \setbox1=\hbox{/} \dimen1=\wd1               
   \ifdim\dimen0>\dimen1                        
      \rlap{\hbox to \dimen0{\hfil/\hfil}}      
      {}#1{}                                    
   \else                                        
      \rlap{\hbox to \dimen1{\hfil${}#1{}$\hfil}}   
   \fi}} 
   
\def \nn {\nonumber}

\title{Large N and Bosonization in Three Dimensions}

\author{Aleksey Cherman and}
\emailAdd{a.cherman@damtp.cam.ac.uk}
\author{Daniele Dorigoni}
\emailAdd{d.dorigoni@damtp.cam.ac.uk}
\affiliation{%
Department of Applied Mathematics and Theoretical Physics,\\ 
University  of Cambridge, \\
Wilberforce Road,
Cambridge CB3 0WA,
UK
}

\abstract{%
Bosonization is normally thought of as a purely two-dimensional phenomenon, and generic field theories with fermions in $D>2$ are not expected be describable by local bosonic actions, except in some special cases.  We point out that 3D $SU(N)$ gauge theories on $\mathbb{R}^{1,1}\times S^{1}_{L}$ with adjoint fermions can be bosonized in the large $N$ limit.  The key feature of such theories is that they enjoy large $N$ volume independence for arbitrary circle size $L$.  A consequence of this is a large $N$ equivalence between these 3D gauge theories and certain 2D gauge theories, which matches a set of correlation functions in the 3D theories to corresponding observables in the 2D theories.   As an example, we focus on a 3D $SU(N)$ gauge theory with one flavor of adjoint Majorana fermions and derive the large-$N$ equivalent 2D gauge theory. The extra dimension is encoded in the color degrees of freedom of the 2D theory.  We then apply the technique of non-Abelian bosonization to the 2D theory to obtain an equivalent local theory written purely in terms of bosonic variables.  Hence the bosonized version of the large $N$ three-dimensional theory turns out to live in two dimensions. 
}

\preprint{DAMTP-2012-57}

\begin{document}

\maketitle
\section{Introduction}
\label{sec:introduction}
The path integral representation of the generating functional for a D-dimensional quantum field theory with fermionic fields $\psi$, bosonic fields $B$, and sources $J$
\begin{align*}
Z[J] = \int \mathcal{D}B \, \mathcal{D}\psi\, \mathcal{D}\bar{\psi} \;\; e^{i \int d^{D}x\, \left(\bar{\psi} F(B,J) \psi  + i \mathcal{L}(B,J) \right)}
\end{align*}
can always be manipulated into a form which only has an explicit dependence on the bosonic fields $B$.  One simply integrates out the fermions 
 giving
\begin{align*}
Z[J] = \int \mathcal{D}B \det( F[B,J] )\;e^{i S(B,J)} = \int \mathcal{D}B\,e^{i\, \rm{tr} \log F(B,J) +i S(B,J)} =\int \mathcal{D}B \,e^{i S_{B}(B,J)},
\end{align*}
which is a description of the theory in terms of a purely bosonic action $S_{B}(B,J)$.  However, the resulting action is in general highly non-local, and this is not what is usually meant by the term `bosonization'.   A bosonized representation of a field theory is one in which a set of new bosonic variables $B'$ are introduced in such a way that the generating functional takes the form
\begin{align*}
Z[J] = \int \mathcal{D}B \,\mathcal{D}B' \,e^{i S'_{B}(B,B',J)} .
\end{align*}
with a \emph{local} action $S'$.  Such bosonized representations are known to be available for field theories living in two dimensions \cite{Coleman:1974bu,Mandelstam:1975hb,Witten:1983ar,DiVecchia:1984df,DiVecchia:1984mf,Gonzales:1984zw}, and are often very useful as they provide a complementary description of the physics.  As just one example, bosonization makes it simple to compute string tensions in a number of 2D gauge theories \cite{Coleman:1974bu,Gross:1995bp}. 

Unfortunately, above two dimensions local bosonized representations of field theories with fermions are not expected to exist in general.   In some cases it is possible to approximately describe the low energy physics of a field theory containing fermion fields using local bosonic actions, as is done in {\it e.g.} chiral perturbation theory for 4D QCD, or for 3D field theories with heavy fermions in \cite{Deser:1988zm,Burgess:1994tm,Fradkin:1994tt,Bralic:1995ip,Banerjee:1995ry,Banerjee:1996qu,LeGuillou:1996dv,LeGuillou:1997zx}.  While for entirely generic higher-dimensional field theories it seems clear that such low-energy bosonization is the most one can hope for, it is natural to ask whether there might be special classes of higher-dimensional field theories for which exact bosonization is possible in the same way as in 2D theories.

Our goal here is to point out that 3D $SU(N_{c})$ gauge theories with $N_{f}$ flavors of adjoint fermions are an example of such a class of theories, and can be bosonized in the large $N_{c}$ limit.  In fact, while our discussion will focus on the 3D example for simplicity, in the conclusions we will argue that our approach should generalize to 4D gauge theories as well.   To find the bosonized description of the 3D gauge theory, we move in two steps.  First, we argue that in the large $N_{c}$ limit, a broad subset of the gauge-invariant correlation functions of the 3D gauge theory can be computed in a certain 2D gauge theory thanks to the phenomenon of large $N_{c}$ orbifold equivalence.    The second step consists of applying the technique of non-Abelian bosonization to the 2D theory, and yields a local two-dimensional bosonic theory which is large-$N_{c}$-equivalent to the original 3D gauge theory with adjoint fermions.

The first step in the argument takes advantage of a remarkable property of large $N_{c}$ gauge theories known in various contexts as volume independence or Eguchi-Kawai reduction \cite{Eguchi:1982nm,Narayanan:2003fc,Kovtun:2007py}. For our purposes, large $N_{c}$ volume independence is the assertion that for a D-dimensional large $N_{c}$ gauge theory compactified on a $D-2$-dimensional spatial torus, so that the theory lives on $\mathbb{R}^{1,1} \times (S^{1}_{L})^{D-2}$ with periodic boundary conditions for the fermions, there is a non-trivial set of correlation functions that are independent of the torus size $L$, up to corrections that are suppressed by powers of $1/N_{c}$, so long as certain symmetry realization conditions are met \cite{Bhanot:1982sh,Kovtun:2007py}. In particular, so long as volume independence holds, correlation functions of a theory on a torus of size $L$ coincide with the correlation functions in the decompactification limit $L\to \infty$. 

For pure Yang-Mills theories, spontaneous breaking of center symmetry invalidates large $N_{c}$ volume independence for torus sizes smaller than a critical size $L_{c}$ set by the strong scale of the theory.  As a result, the small-volume physics is volume-dependent and very different from the physics seen for  $L> L_{c}$, where the large $N_{c}$ theory becomes $L$-independent.   However, it was recently realized that the situation is very different for theories with adjoint fermions with periodic boundary conditions on the torus, because the adjoint fermions prevent the center-symmetry-breaking phase transition from taking place.  As a result, YM theories with adjoint fermions enjoy the property of large $N_{c}$ volume independence for \emph{arbitrarily} small $L$ \cite{Kovtun:2007py}.

The existence of general recipes for bosonization of 2D theories and the phenomenon of large $N_{c}$ volume independence for arbitrarily small $L$ in the theories we consider suggests a natural strategy for obtaining bosonized descriptions of $SU(N_{c})$ YM theories with adjoint fermions.  We  send the spatial $D-2$-dimensional torus size to zero, and perform a dimensional reduction in a way consistent with large $N_{c}$ volume independence, obtaining a two-dimensional field theory which at large $N_{c}$ is equivalent (in a precise sense to be defined below) to the original $D$-dimensional field theory.    The application of non-Abelian bosonization \cite{Witten:1983ar} to the 2D theory then produces a local bosonic theory which has correlation functions that coincide with those of the original 3D fermionic theory in the large $N_{c}$ limit.   

We hasten to emphasize that the dimensional reduction procedure relevant for such theories is \emph{not} the naive one where one simply truncates the theory to the lowest Kaluza-Klein modes, which would of course have no chance of producing a theory equivalent to the original large-$L$ theory.  Instead, as we explain in Section~\ref{sec:Orbifolding}, the systematic way to perform dimensional reduction for such theories uses the machinery of orbifold projections \cite{Kovtun:2007py},  involves the use of gauge-invariant lattice regularization for modes on $S^{1}$, and results in 2D theories with unitary scalar fields rather than the Hermitian fields which would arise from naive dimensional reductions.

In this paper we focus on the simplest example of these ideas, which involves studying $3D$ $SU(N_{c})$ gauge theory with one flavor of adjoint Majorana fermions with bare quark mass $m$ on $\mathbb{R}^{1,1} \times S^{1}_{L}$, which is described by the action
\be
S_{3D} = \int{d^{3}x\,  \tr \left( -\frac{1}{2g_{3}^{2}} F_{\alpha \beta}F^{\alpha \beta} + \bar{\psi} \left[i \slashchar{D}  - M\right] \psi \right)} \; ,
\ee
where $g_{3}$ is the Yang-Mills coupling, $\psi$ is a Majorana fermion field and $D_{\alpha} = \partial_{\alpha} - i  [A_{\alpha}, \cdot]$ with $A_{\alpha} = A_{\alpha}^{i} t_{i}$ the $SU(N_{c})$ gauge field with field strength $F_{\alpha \beta}$, $t_{i}, i=1,\ldots, N_{c}^{2}-1$ are the Hermitian generators of $SU(N_{c})$ obeying $\tr t_{i} t_{j} = \frac{1}{2} \delta_{ij}$, $\alpha, \beta = 0,1,2$, and we use the mostly-minus metric convention throughout.  In the limit $M \to 0$, this theory becomes $\mathcal{N}=1$ super-Yang-Mills theory.  The bosonized theory is obtained from the 3D theory on a discretized $S^{1}$ direction with a $\Gamma$-site lattice spacing $a$, $L=\Gamma a$. We show that in the large $N_{c}$ limit,  the physics described by Equation~\eqref{eq:3DTheory} can also be described by the action
\begin{align}
\label{eq:2DBosonizedTheory}
 S= &\int d^2x \left\lbrace \,\mbox{Tr}\left(\frac{-a}{2g_{3}^2} F_{\mu \nu}^2+
\frac{1}{a g_{3}^2} \vert D_\mu \phi\vert^2 + \frac{N}{8\pi} \vert D_\mu g\vert^2
\right)+
\tilde{m}^{2} \,\mbox{Tr} \,g\, \mbox{Tr}\, g^\dagger - \frac{\tilde{c}}{a^{2}} \mbox{Tr} (g \phi) \mbox{Tr} (g^\dagger \phi^\dagger)
\right\rbrace \nonumber\\
&+N \tilde{\Gamma} (g,A)\,,
\end{align}
where $\mu,  \nu = 0,1$, $g$ encodes the bosonized representation of the fermions and is matrix-valued field living in the group $SU(N_{c})/\mathbb{Z}_{N_{c}}$, with $\mathbb{Z}_{N_{c}}$ being the center subgroup of $SU(N_{c})$, $\phi \in SU(N_{c})$ and is related to the component of the original gauge field in the $S^{1}$ direction, $\tilde{m}$ is a mass parameter related to the original quark mass $M$,  and the dimensionless parameter $\tilde{c}$ is discussed in Section~\ref{sec:DimensionalReduction} and Section~\ref{sec:3DBosonization} and has to do with the discretization we adopt.  Finally, the second line contains an appropriately gauged Wess-Zumino-Witten term.

We hope to make this paper mostly self-contained, and hence organize the exposition as follows.  In Section~\ref{sec:DimensionalReduction} we  explain the orbifold projections relating the original 3D theory and its large $N_{c}$-equivalent 2D form, using the technology developed in \cite{Bershadsky:1998cb,Schmaltz:1998bg,Kovtun:2003hr,Kovtun:2004bz,Kovtun:2007py}.  On a first encounter the phenomena of large $N_{c}$ volume independence and large $N_{c}$ equivalences between 2D and 3D gauge theories look quite counterintuitive, and so in Section~\ref{sec:SpectrumVsVacuum} we work through a calculation of the vacuum energy in the 2D theory in perturbation theory, which  crisply illustrates the way in which the extra dimension is encoded in the color degrees of freedom of the 2D theory.  The calculation serves as a perturbative verification that the vacuum energy of the 2D theory vanishes so long as the fermions are massless, as it must, since at $m=0$ the 3D theory has $\mathcal{N}=1$ supersymmetry and must have vanishing vacuum energy.   

Next, in Section~\ref{sec:AdjointBosonization} we briefly review the derivation of the technique of  non-Abelian bosonization of adjoint fermions, focusing on the points which present particular subtleties for our application.   Finally,  in Section~\ref{sec:3DBosonization} we use the previous results to bosonize the 2D theory which is large-$N_{c}$ equivalent to 3D YM with adjoint fermions, obtaining a theory with the action given in Equation~\eqref{eq:2DBosonizedTheory}.  
In Section~\ref{sec:Conclusions} we summarize the results and outline some promising directions for generalizations and applications.

\emph{Note Added:}  As this paper was being finalized, a very interesting preprint by Aharony, Gur-Ari, and Yacobi appeared proposing an (apparently) unrelated approach to the bosonization of 3D theories of free fermions \cite{Aharony:2012nh}.

\section{Dimensional reduction of a volume-independent theory}
\label{sec:DimensionalReduction}

The starting point of our analysis is establishing the relation between the YM theory with adjoint fermions theory living on $\mathbb{R}^{1,1} \times S^{1}_{L}$ and a large-$N_{c}$-equivalent theory living on $\mathbb{R}^{1,1}$.   The 3D theory has the action in Equation~\eqref{eq:3DTheory}:
\be
\label{eq:3DTheory}
S_{3D} = \int_{\mathbb{R}^{1,1} \times S^{1}_{L}}{d^{3}x\,  \tr \left( -\frac{1}{2g_{3}^{2}} F_{\alpha \beta}F^{\alpha \beta} + \bar{\psi} \left[i \slashchar{D}  - M\right] \psi \right)} \; .
\ee
with $\alpha, \beta = 0,1,2$, with the condition $x_{2} \sim x_{2}+L$ and periodic boundary conditions for all fields on the $S^{1}$, our convention for gamma matrices in 3D is $\gamma^0= \sigma_1,\gamma^1=i\sigma_2, \gamma^2 = i\sigma_3$, and $\psi$ is a two-component Majorana spinor.  With our conventions, the Majorana condition on the spinor components of $\psi$ is given by $\psi=(\psi_1,i\psi_2)$ with both $\psi_i^\ast=\psi_i$.   In the limit $M\to 0$, the action becomes that of $\mathcal{N}=1$ super-Yang-Mills theory, but it is conjectured that the supersymmetry is spontaneously broken unless one turns on a Chern-Simons term with a coefficient which is larger than $N_{c}/2$~\cite{Witten:1999ds}~\footnote{For an interesting recent analysis of 3D $\mathcal{N}=1$ SYM theory, see \cite{Agarwal:2012bn}, which is based on the approach developed in \cite{Karabali:1995ps,Karabali:1996je,Karabali:1997wk,Karabali:1998yq,Agarwal:2007ns}.}.

Aside from the obvious spacetime symmetries, the theory has an $SU(N_{c})$ gauge symmetry, and also a global symmetry known as center symmetry.  Center symmetry plays a very prominent role in the lattice gauge theory and continuum studies of confinement (see e.g. \cite{Greensite:2003bk,Unsal:2008ch}), and appears also in the context of string theory, see e.g. \cite{Aharony:1998qu,Poppitz:2010bt}.  This symmetry is somewhat unusual in that its existence as a non-trivial symmetry depends on the topology of the spacetime on which the field theory lives.  To recall how such a symmetry arises, note that constant gauge transformations that differ by an element of the center subgroup $\cong \mathbb{Z}_{N_{c}}$ of $SU(N_{c})$ act trivially on all of the fields in the theory, so that one really has an $SU(N_{c})/Z_{N_{c}}$ symmetry acting on the fields \cite{Witten:1978ka}.   Integrating over the gauge fields amounts to taking a quotient by $SU(N_{c})$, and the quotient $Z = [SU(N_{c})/\mathbb{Z}_{N_{c}} ] /SU(N_{c}) \cong \mathbb{Z}_{N_{c}}$ is the global symmetry, known as the center symmetry.  If the theory had lived in $\mathbb{R}^{1,2}$, all observables in the theory would transform trivially under $Z$, and one would not really regard it as a symmetry worthy of the name.  However, on $\mathbb{R}^{1,1} \times S^{1}_{L}$ the situation is different because Wilson lines that wrap the $S^{1}$ transform non-trivially under the action of $Z$, which acts by changing the periodicity condition on $A_{2}$ by multiples of elements of the $SU(N_{c})$ group center.  Hence Wilson loops wrapping $S^{1}$ transform as
\begin{align}
\langle \tr \mathcal{P} e^{i \int_{S_{1}}dx_2\, A_{2}(x^{\mu},x_{2})} \rangle = \langle \Omega(x^{\mu}) \rangle  \to \omega \langle \Omega(x^{\mu}) \rangle
\end{align}
where $\mathcal{P}$ represents path ordering and $\omega \in \mathbb{Z}_{N_{c}}$. So $Z$ is a bona-fide global symmetry of the theory.

The theory defined by Equation~\eqref{eq:3DTheory} has distinct phases depending on whether the ground state is invariant under the center symmetry, with the phases distinguished by the expectation values of $\tr \Omega^{n}, |n| < N_{c}$.   Phases of a theory with unbroken center symmetry are known as  `confining', with a linear rising potential between widely separated fundamental color test sources, while phases where the center symmetry is broken are known as deconfined phases \cite{Greensite:2003bk}.  Three-dimensional YM with one flavor of adjoint Majorana fermions is expected to be confining in large volumes $\lambda_{3} L = g_{3}^{2} N_{c} L \gg 1$, with an unbroken center symmetry, which is also known to be true of pure YM theory.  The remarkable feature of YM theories with adjoint fermions is that they remain confining even at small volumes $\lambda_{3} L \ll 1$, as can be seen from a perturbative calculation of the effective potential for the eigenvalues of a Wilson loop wrapping the $S^{1}$ direction \cite{Kovtun:2007py}.  Such calculations show that so long as the fermions have periodic boundary conditions on $S^{1}$, so that one is dealing with a spatial compactification as opposed to a thermal one, the effective potential is minimized on configurations where the $N_{c}$ eigenvalues of $\Omega$ are evenly spaced on the unit circle. This implies 
\begin{align}
\label{eq:CenterSymmetricVacuum}
\langle \tr \Omega^{n} \rangle = 0 \qquad \textrm{for all}\,\, n\neq0 \in \mathbb{Z}, |n|<N_{c}
\end{align}
and hence center symmetry is preserved.

We now explain the construction of the 2D theory which is large-$N_{c}$ equivalent to the 3D theory.  As we will see, it is not given by a conventional dimensional reduction of the 3D theory.  The most naive dimensional reduction for small $L$ would truncate Equation~\eqref{eq:3DTheory} to the Kaluza-Klein zero-modes, yielding
\begin{align*}
S_{\rm naive} = L\int_{\mathbb{R}^{1,1} }{d^{2}x\, \tr \left[ -\frac{1}{2g_{3}^2} F_{\mu \nu}^{2} +  \frac{1}{g_{3}^2} \tr |D_{\mu} H|^{2} 
+ \bar{\psi} (i  \slashchar{D}+i\gamma_\ast[H,\,\cdot \,]  - M) \psi \right]} \; .
\end{align*}
with $\mu, \nu = 0,1$, $\gamma_{\ast} \equiv i\gamma_{0} \gamma_{1} = -\gamma_{2}$ is the chiral gamma matrix in 2D, and $H$ is the KK zero mode of the $A_{2}$ component of the gauge field.  Such a reduction is \emph{wrong} for 3D YM with adjoint fermions, since it amounts to an expansion around a phase with broken center symmetry $\langle A_{2}  \rangle =0$, and hence $\langle H \rangle = 0$. This is not the correct phase for YM with adjoint fermions either at large $L$ or small $L$.    

A less naive dimensional reduction, which would be consistent with the global symmetry of the 3D theory, would consist of an expansion around a vacuum where $A_2$ (and hence $H$) has a vacuum expectation value (VEV) consistent with center symmetry, for instance the Hermitian traceless configuration, $\langle H \rangle  ={ \rm diag}\,(2\pi\,(-N_c+1)/N_c,2\pi\, (-N_c+3)/N_c, \ldots, 2\pi(N_{c}-1)/2N_{c})$, leading to a breaking of the $SU(N_{c})$ gauge symmetry down to $U(1)^{N_{c}-1}$.
For our purposes, such an approach suffers from two	 problems, one conceptual and one practical.

The conceptual problem is that expanding around any particular such VEV requires gauge fixing, since the values of the entries of the matrix $A_{2}$ are gauge-dependent.  Gauge fixing would be an inauspicious way to start an analysis which is meant to work beyond perturbation theory, thanks to notoriously tricky issues such as Gribov ambiguities.  Since our aim is to find a dimensional reduction which would make sense outside of perturbation theory, this is a strong reason to avoid such an approach.

There is also a practical problem with doing a dimensional reduction with a non-trivial VEV for $A_{2}$.  Once we fix the VEV in Equation~\eqref{eq:CenterSymmetricVacuum} for the Polyakov loop, the adjoint Higgs mechanism gives a theory where the KK zero modes are separated from the non-zero modes by energies of order $1/(N_{c}L)$, not $1/L$ as one might have naively expected.  The factor of $1/N_{c}$ in the spacing of modes carrying KK momenta arises from the spacing of the eigenvalues of $H$ demanded by center symmetry, and indeed is the heuristic reason for the volume independence of gauge theories in center-symmetric phases.  In the `t Hooft large $N_{c}$ limit where $\lambda_{3D} L$ is fixed as $N_{c} \to \infty$, which is the most interesting one from the perspective of the 3D theory, one would have to keep the whole infinite tower of such states, and the resulting `2D' theory would not truly look two-dimensional.

To construct the desired 2D theory, we need an approach in which a manifestly gauge-invariant dimensional reduction can be defined, while allowing for the preservation of center symmetry.    It turns out that the necessary approach is to first discretize the $S^{1}$ direction on a lattice with $\Gamma$ sites and lattice spacing $a$, so that $L= \Gamma a$, and borrow the standard technology of lattice gauge theory to keep the resulting theory manifestly gauge invariant.  Then the volume-independence of the theory translates into the independence of the theory on the number of lattice sites $\Gamma$.  With such a formulation, there is a very natural and simple way to define dimensional reduction to 2D that does not require gauge fixing or a commitment to expanding around the wrong vacuum.   One simply defines the desired 2D theory to be given by discarding all of the lattice sites except for one, which removes all link fields except for one, yielding a theory living on $\mathbb{R}^{1,1}$~\cite{Bedaque:2009md}.  Of course, one must then explain why such a prescription would yield a theory which is equivalent at large $N_{c}$ to the original 3D theory.  The reason for the equivalence is because such a 2D theory is connected to the 3D theory by orbifold projections which imply a large $N_{c}$ equivalence \cite{Kovtun:2007py}.

Before explaining the orbifold equivalence, we write down the discretized theory. Borrowing language from the story of dimensional deconstruction \cite{ArkaniHamed:2001ca}, the discretized theory is defined by encoding $A_{2,n} \equiv A_{2}(x_{2} = n a)$ in unitary link fields $U_{n} = e^{i a A_{2,n}}, n = 1, \ldots, \Gamma$, and the Lagrangian takes on form of a quiver gauge theory with $\Gamma$ $SU(N_{c})$ nodes with $\phi_{n}$ transforming as $(\overline{F}, F)$ under $[SU(N_{c})_{n-1}, SU(N_{c})_{n}]$.    The fermion discretization is more subtle and described below.  Our discretization of the 3D theory is defined by the Lagrangian
\begin{align}
\label{eq:3DLat}
S_{\rm lat} &= \int d^{2}x\, a \, \sum_{i=1}^{\Gamma}\tr \left. \Bigg[ -\frac{1}{2g_{3}^2} (F^{i})_{\mu \nu}F^{i,\mu \nu} + \frac{1}{a^2 g_{3}^{2}} |D^{i}_{\mu} \phi_{i}|^{2} \right. \nn \\
&\left.+\bar{\psi}_{i} (i  \gamma^{\mu} D_{\mu}^{i}  - m) \psi_{i} + \frac{i c }{2a} \bar{\psi}_{i} \gamma_{3}\left\{\phi_{i} \psi_{i+1} \phi_{i}^{\dag} - \phi_{i-1}^{\dag} \psi_{i-1} \phi_{i-1}\right\}  \right. \\
&\left.+ \frac{r c}{2a} \bar{\psi}_{i} \left\{ \phi_{i} \psi_{i+1} \phi_{i}^{\dag} + \phi_{i-1}^{\dag} \psi_{i-1} \phi_{i-1} - 2 \psi_{i}\right\} + m \bar{\psi}_{i} \psi_{i} \right. \Bigg] \; \nn.
\end{align}
where $D^{i}_{\mu} \phi_{i}= \partial_{\mu} + i A^{i}_{\mu} \phi_{i} - i \phi_{i}A^{i+1}_{\mu} , D^{i}_{\mu} \psi =  \partial_{\mu} \psi + i [A^{i}_{\mu}, \psi]$.  

The fermion terms and the parameters $m, r$ and $c$ deserve explanation.  On the second line, there is a standard fermion kinetic term for 2D fermions, and a term which becomes the kinetic term along $x_{2}$ in the continuum limit $\lambda_{3D} a\to 0$.  As is well known, however, naive discretizations of fermions suffer from fermion doubling, meaning that with only the terms in the second line we would end up with a continuum theory with two flavors of fermions, not one.   

We deal with this in the simplest and least sophisticated way, by including a Wilson term with coefficient $r$ in the third line.  Eventually, we will set $r=1$ as is standard in the lattice literature.  The Wilson term, which in the continuum limit approaches $a \bar{\psi} D^{2}_2 \psi$, serves to break the lattice doubler symmetry \cite{montvay1997quantum} which ensures fermion doubling for the naive fermion action with $r=0$.  The Wilson term gives the doubler fermions masses of order $1/a$, so that they decouple in the continuum limit.  However, the Wilson term also breaks parity symmetry (the analog of chiral symmetry in 4D), and as a result the physical fermions also pick up an additive mass shift of order $1/a$ thanks to interactions.  To get a single flavor of fermions with an adjustable mass, we add a fermion mass term $m$, which has to be tuned to produce the desired `physical' mass, and $m \neq M$.   

Another subtlety we must deal with has to do with renormalization of the discretized fermion kinetic term, and this is the reason the Wilson kinetic term appears in Equation~\eqref{eq:3DLat} multiplied by the dimensionless parameter $c$.  The point is that quantum corrections will renormalize $m$, $c$, and the coefficient of $|D\phi|^{2}$.  Above we have already discussed the renormalization of $m$,  and the renormalization of the coefficient of $|D\phi|^{2}$ can be viewed as simply a renormalization of the value of $a$.   To make sure the fermions see the same `speed of light' in the $S_{1}$ direction as the gauge bosons, however,  one must tune 
$c$ appropriately, as is well-known in the literature on anisotropic lattice actions, see {\it e.g.} \cite{Karsch:1982ve,Burgers:1987mb, Klassen:1998ua,Klassen:1998fh,Chen:2000ej,Harada:2001ei,Bedaque:2007xg}.  At tree level in perturbation theory, the correct value of $c$ to get a Lorentz-invariant continuum limit would be $c=1$, but in general $c \neq 1$ to get the continuum limit we want.  Tuning $c$ gives us control over the only marginal operator\footnote{The fact that one does not need to introduce separate $c$ parameters for each term in the sum over lattice sites is due to the discrete lattice translation symmetry.} consistent with the symmetries which could break Lorentz invariance, and hence will be enough to give a Lorentz-invariant theory in the continuum limit\footnote{We are grateful to Francis Bursa, Tom Hammant, and Matt Wingate for very useful discussions about these issues.}.

Finally, we note that while in 4D gauge theories the tuning of parameters analogous to $m$ and $c$ necessary to get the desired continuum limit is in general non-perturbative, the situation is better in 3D gauge theories, since 3D YM theory is super-renormalizable \cite{Reisz:1987da,Golterman:1988ta,Catterall:2000rv,Kaplan:2003uh,Giedt:2004vb,Elliott:2005bd,Elliott:2007bt}.  Consequently it is generally possible to determine the necessary tuning of $m, c$ by finite-order analytically calculations in lattice perturbation theory, and doing so for the particular theory we are considering here is an interesting direction for future work.

\subsection{Orbifold projection}
\label{sec:Orbifolding}

Having explained the discretization of the $S^{1}$ direction of the 3D theory, we are finally in a position to define the dimensional reduction down to two dimensions and to explain the reason for the large $N_{c}$ equivalence between the 3D and 2D theories.

The point is that discarding all lattice sites except for one can be viewed as an orbifold projection.  For our purposes, an orbifold projection is a quotient of a `mother' theory by some discrete symmetry $G$, which defines a daughter theory.  In the case of interest to us, it is also possible to define an orbifold projection of the daughter theory that yields back the mother theory, but with a reduced number of colors.  Remarkably, for theories linked by orbifold projections, it has been shown that the planar diagram expansions of the theories coincide to all orders in perturbation theory\footnote{Provided that the action of the daughter theory is rescaled by a factor of $\rm{rank}(G)^{-1}$ before one tries to compare the surviving correlation functions.}, leading to a large $N_{c}$ equivalence between correlation functions which survive the projection \cite{Bershadsky:1998cb,Schmaltz:1998bg}.  In fact,  in \cite{Kovtun:2003hr,Kovtun:2004bz} it was proved that for theories with matter in two-index representations,  large $N_{c}$ orbifold equivalence holds non-perturbatively provided the symmetries used in these projections are not spontaneously broken.

To apply the orbifold projection/equivalence machinery to our case, note that the discretized 3D theory has a discrete translation group $T_{S^{1}} \cong \mathbb{Z}_{\Gamma}$.  The orbifold projection to the single-site 2D theory is simply a projection based on the  $\mathbb{Z}_{\Gamma}$ translation symmetry\footnote{Reduction to theories with more than one site would involve orbifold projections based on non-trivial subgroups of the discrete translation group}.   Such an orbifold projection discards all of the nodes of the quiver except for one, leaving only one link field, which becomes an adjoint unitary scalar.  The action of the 2D daughter theory obtained from the orbifold projection takes the form
\begin{align}
S_{2D} &\notag =  \int{d^{2}x\, \tr\left(-\frac{1}{2g_2^2} F_{\mu\nu}F^{\mu \nu} + \frac{1}{ a^2 g_2^2} |D_{\mu} \phi |^2+
 \bar{\psi} \left( i \slashchar{D} -m \right)\psi + \right. }\\
& \label{eq:2DTheory} \left.  - \frac{c}{2a} [\bar{\psi}, \phi] \gamma_{*} \{ \psi, \phi^{\dagger}\} +
 \frac{r c}{2a} [\bar{\psi}, \phi] [ \psi, \phi^{\dagger}] \right)
\end{align}
where $g_2$ is the two-dimensional gauge coupling related to $g_{3}$ via $g_2^2=g_3^2/a $, $\phi$ is a unitary scalar living in $SU(N_{c})$ and is the remaining link field, $\alpha = 0,1$,  $\gamma_{*} = i \gamma_{0} \gamma_{1}=\sigma_3$, the fermion fields have been rescaled by a factor of $a$ to make sure they are canonically normalized in 2D, and we have included the rescaling of the projected action by a factor of $1/\Gamma$ necessary for the 2D theory to have the same correlation functions as the original 3D theory in the large $N_{c}$ limit. The first fermionic term is simply the usual kinetic term for an adjoint Majorana fermion with bare mass $m$ in two dimensions.  The second term is the remnant of the fermion kinetic term in the $S^{1}$ direction of the 3D theory, while the third one is the remnant of a Wilson term in the 3D theory. 

The 2D theory defined by eq. \eqref{eq:2DTheory} has a $SU(N_{c})$ gauge symmetry, with $\psi$ and $\phi$ transforming in the adjoint representation.  There is a $G_{c} = \mathbb{Z}_{N_{c}}$ global symmetry that acts as $\phi \to \omega \phi$ with $\omega \in \mathbb{Z}_{\Gamma}$, with the other fields transforming trivially.  This is the remnant of the center symmetry of the 3D theory\footnote{Provided $N_{c} = N \Gamma$ for some integer $N$, which can certainly be arranged for any desired $\Gamma$ if $N_{c}$ is large, one can define an orbifold projection based on a $Z_{\Gamma}$ subgroup of $G_{c}$ and its embedding into $SU(N_{c})$ in such a way that the resulting daughter theory is again a discretized 3D theory on $\Gamma$ sites, but now with $N$ colors \cite{Kovtun:2007py,Bedaque:2009md}.}.  There is also a discrete symmetry acting as either $\psi_{L} \to -\psi_{L}$ or $\psi_{R} \to -\psi_{R}$, provided that at the same time one sends $\phi \to \phi^{\dag}, m\to-m$ and $r\to-r$, which will be very useful in Section~\ref{sec:3DBosonization}.

Large $N_{c}$ orbifold equivalence implies that the 2D theory will be equivalent to the original 3D theory only so long as discrete translation symmetry in the 3D theory is unbroken and the center symmetry of the 2D theory, which is probed by the order parameters $\langle \mbox{Tr}\phi^{n} \rangle$ is unbroken \cite{Bhanot:1982sh,Kovtun:2007py}.  The first condition is not expected to be violated in gauge theories without chemical potentials.  Given the effective potential calculations in theories with adjoint fermions cited above \cite{Kovtun:2007py}, one might suppose that the second condition is sure to be met thanks to the presence of adjoint fermions in the theory.  However, the realization of center symmetry in the 2D theory is somewhat subtle \cite{Bedaque:2009md,Bringoltz:2009mi,Bringoltz:2009fj,Poppitz:2009fm}.  

The 2D theory is a gauged non-linear sigma model.  Non-linear sigma models without gauge fields are renormalizable in two dimensions (and indeed can be exactly solved \cite{Polyakov:1983tt,Wiegmann:1984pw,Fateev:1994ai}), and of course gauging does not break renormalizability.   However, the gauging breaks the $SU(N_{c})_{L} \times SU(N_{c})_{R}$ symmetry of the sigma model, which acts as $\phi \to L \phi R^{\dag}$, down to the diagonal $SU(N_{c})$ subgroup $L=R$.  As a result one can add a very large number of marginal terms to the Lagrangian in Eq.~\eqref{eq:2DTheory} consistent with its remaining symmetries\footnote{There is an analogue of this in the original 3D theory, to which we could add a (non-local) potential $V = L^{-4} \sum_{n} c_{n} \tr |\Omega^{n}|^{2} + \ldots$ which is consistent with all of the symmetries of the theory but may affect the realization of the center symmetry of the theory.   These are simply the double-trace deformations introduced in \cite{Myers:2007vc,Ogilvie:2007tj,Unsal:2008ch}.}, namely
\begin{align}
V = \frac{1}{a^{2}} \left( \sum_{n=1} c_{n} |\tr \phi^{n}|^{2} +  \frac{1}{N_{c}}\sum_{n,m=1} d_{n,m} \tr \phi^{n} \tr \phi^{m} \tr (\phi^{\dag})^{-(n+m)} + \ldots \right)
\end{align}
Clearly, these terms amount to a potential for the eigenvalues of $\phi$, and hence by tuning these coefficients one can affect the realization of the center symmetry.  Whether or not such terms are added to the original action, they will be generated by quantum fluctuations, which will lead to a renormalization of the values of $c_{n}, d_{n,m}, \ldots$.  If one chooses not to add the above ``double-trace'' potential to the action and lets one be generated just from quantum fluctuations, the result will obviously be regularization scheme dependent.  Amusingly, as shown by \cite{Bringoltz:2009mi}, a complete discretization of the 2D theory on a lattice with the same spacing in all directions and a Wilson fermion action with no tree-level double-trace potential produces an effective potential of the form $V$ with coefficients that drive the theory to a center-symmetric vacuum.  If one uses some other regularization that does not share this fortunate feature, one could still tune the coefficients in $V$ to arrange for the center symmetry to be preserved \cite{Poppitz:2009fm}.  Hence large $N_{c}$ orbifold equivalence can be arranged to be valid for the 2D theory we will be studying, and in what follows we will assume that this has been done.

The claim that Equation~\eqref{eq:3DTheory} and Equation~\eqref{eq:2DTheory}, possibly deformed by the addition of $V$, define equivalent theories at large $N_{c}$ may raise a number of natural questions, which we should address before discussing bosonization.  
\begin{enumerate}
\item If the 2D theory is supposed to be equivalent to the 3D theory at large $N_{c}$, where is the extra dimension encoded in the 2D theory?  
\item Does the 2D theory know about the fact that in the continuum limit the 3D theory has $\mathcal{N}=1$ supersymmetry if $m=0$, with powerful consequences for e.g. the vacuum energy of the theory?  
\item In the original 3D theory, at least for large $L$, there is certainly no need to add any potential of the form of $V$ to the action. But if the 2D and 3D theories are to be equivalent, and a non-trivial $V$ needs to be included in the definition of the 2D theory, it better be the case that the common-sector physics in the 2D theory does not depend on the values of the parameters in $V$ so long the theory is in a center-preserving phase. But how does that work?  
\end{enumerate}

In the following section we will investigate the second point by a study of the spectrum of our theory at the leading order in perturbation theory around the center-symmetric vacuum.  This will allow to see the fate of supersymmetry by calculating the vacuum energy in the 2D theory in perturbation theory.  As it happens, the calculation also serves to illustrate the answers to the other two questions in a particularly simple context.  (Of course, the answers to the first and third questions have already been addressed in the seminal works on volume independence \cite{Eguchi:1982nm,Kovtun:2007py,Unsal:2008ch}.)

\section{Spectrum of modes and the vacuum energy}
\label{sec:SpectrumVsVacuum}
\subsection{Perturbation theory and the spectrum}
\label{sec:PerturbationTheory}

To illustrate how the dimensional reduction works, in this section we explore its properties in the simplest possible context.  Specifically, we will compute the spectrum of our 2D theory in perturbation theory, and compare to expectations from 3D.   The technical goal of the analysis is to verify that in an appropriate limit of the 2D theory corresponding to the emergence of $\mathcal{N}=1$ supersymmetry in the 3D theory, the vacuum energy vanishes.  

An immediate problem we must face before beginning the calculation is that as far as observables like the spectrum is concerned, in the large $N_{c}$ limit with the 't Hooft coupling $\lambda_{3} = g_{3}^{2}N_{c}$ and circle size $L$ held fixed (more precisely holding the dimensionless quantity  $\lambda_{3} L$ fixed), the theory is necessarily in a strongly interacting regime.  After all, as long as continuum volume independence holds, which it does in such a limit, the physics at any $L$ will be large $N_{c}$ equivalent to the physics for $L\to \infty$, where we expect the spectrum to be inaccessible to weak coupling calculations.

However, it is possible to find a regime where the large $N_{c}$ 3D theory becomes weakly coupled.  The trick is to consider a non-'t Hooft large $N_{c}$ limit with $L\sim N_{c}^{-1}$, so that $\lambda_{3} L N_{c}$ is fixed as $N_{c}\to \infty$.  The downside of such a limit is that it brings us outside the regime of validity of continuum volume independence, since the properties of the resulting theory \emph{will} depend on $L$~\cite{Unsal:2008eg}.   The positive aspects of considering such a regime are that 
\begin{enumerate}
\item If $\lambda_{3} L N_{c} \ll 1$ one expects perturbation theory and semiclassical approximations to become reliable guides to the properties of the theory.  
\item Center symmetry is expected to be preserved for any $\lambda_{3}L N_{c}$ (as can be checked explicitly if $\lambda_{3}L N_{c} \ll 1$), and the theory will still be in a confining phase\footnote{See \cite{Shifman:2008ja,Unsal:2008eg,Shifman:2009tp} for discussions of how this happens in the parallel 4D case.}. 
\end{enumerate}

This suggests that if we do a perturbative calculation in the 2D theory and take large $N_{c}$ equivalence to the 3D theory for granted, we should only expect the calculation to be controlled for  $\lambda_{3} L N_{c} \ll 1$~\footnote{For other recent perturbative studies of large $N_{c}$ gauge theories in center-symmetric phases, see~\cite{Bringoltz:2009mi,Bringoltz:2009fj,Poppitz:2009fm,Hollowood:2009sy,Myers:2007vc,Myers:2008zm,Myers:2009df,Ogilvie:2007tj}.}.  However, once the 3D theory is out of the volume-independent regime, it less heuristically obvious that the large $N_{c}$ limit of the 2D theory should continue to capture the 3D physics.  In particular, the non-perturbative arguments for large $N_{c}$ equivalence presented in \cite{Kovtun:2003hr,Kovtun:2004bz} were constructed for 't Hooft large $N_{c}$ limits, not the large $N_{c}$ limit we are driven to consider.  However, the perturbative arguments of \cite{Bershadsky:1998cb,Schmaltz:1998bg}, which match the planar Feynman diagrams between the two theories, will still apply.  So at least in perturbation theory, the dimensionally reduced 2D theory we have constructed will still be large $N_{c}$ equivalent to the 3D lattice theory (and hence the continuum limit of the 3D theory as well) even when $\lambda_{3} L N_{c} \ll 1$.  Our calculations in this section will be perturbative, and so this is good enough for our present purposes.  It is tempting to speculate that large-$N_{c}$ equivalence for $\lambda_{3} L N_{c} \ll 1$ will hold non-perturbatively as well, but an exploration of this is beyond the scope of the current investigation. 

With this subtlety addressed, we proceed to the calculation of the spectrum in perturbation theory\footnote{A pioneering example of such calculations in center-symmetric vacua, see \cite{GonzalezArroyo:1982hz}.}.  As in any perturbative calculation, we must start by choosing an appropriate gauge and vacuum to expand around.   As we have been stressing, the presence of adjoint fermions and an appropriate choice of $V$ will guarantee that the reduced theory vacuum will preserve the center symmetry.  For our perturbative calculation, we choose a gauge in which the unitary scalar $\phi$ takes the VEV
\begin{equation}
 \langle\phi\rangle=\left(
\begin{matrix}
 1 & 0 & \cdots  & 0\\
0 & \omega & 0  & \vdots\\
\vdots & 0 & \ddots  & 0\\
0 & \cdots &0 & \omega^{N-1}\\
\end{matrix}
\right) \equiv \Omega\,.
\end{equation}
We will determine the spectrum of the theory to leading order in perturbation theory by expanding $\phi$ around this vev:
\begin{align}
\phi =& \Omega \, e^{iH} \simeq \Omega \left(1+iH +O(H^2) \right)\,, 
\end{align}
where we introduced the Hermitian matrix $H$, which is assumed to have zero expectation value.

The covariant derivative for $\phi$ introduces some mixing terms between $A_\mu$ and $H$.  These mixing terms can be removed by working in $R_{\xi}$ gauge, which amounts to adding an appropriate  gauge-fixing term and ghosts to the Lagrangian. The gauge-fixing function and Lagrangian are
\begin{align}
 G=&\frac{1}{\sqrt{\xi}} \left[ \frac{\partial_\mu A^\mu}{g_2}- \frac{\xi }{a^2 g_2} \left(\Omega H \Omega^\dagger-H\right)\right]\,,\;\;
\mathcal{L}_{\rm gf}= - \mbox{Tr} \left(G^2\right)\,.
\end{align}
Under infinitesimal gauge transformations parametrized by $\alpha = \alpha^{i}t_{i}$, the fields appearing in $G$ transform as
\begin{align}
 \delta H =& - \left(\Omega^\dagger \alpha \,\Omega -\alpha\right) (1+iH)-i[\alpha,H]\,, \;\;
\delta A_\mu = D_\mu \alpha \,,
\end{align}
and from here it is straightforward to compute $\delta G / \delta \alpha$, obtaining the ghost action 
\begin{equation}
 \mathcal{L}_{\rm gh} = -\mbox{Tr} \left[
\frac{1}{g_2}\bar{c} \,\partial_\mu D^\mu c-\frac{\xi}{a^2 g_2}\left(
\bar{c}\, \Omega^\dagger c\, \Omega+\bar{c}\, \Omega\, c\, \Omega^\dagger -2\bar{c} c
\right)
\right]\,.
\end{equation}

To read off the spectrum of the theory to the leading order in perturbation theory we simply need the quadratic part of the action.  The interaction terms will make small contributions so long as $\lambda_{3} L N_{c} \ll 1$.   After canonically normalizing all of the fields, the quadratic part of the action can be written
\begin{align}
\label{eq:QuadraticAction}
 S_{\rm quad}=\int d^2 x\,& 
\left\lbrace A_{\mu\,nm}
 \left[ \square \,g^{\mu\nu}-
 \left( 1-\frac{1}{\xi}\right)
 \partial^\mu\partial^\nu+
\frac{4}{a^2} \sin^2\left(\frac{\pi(n-m)}{N}\right)g^{\mu\nu} \right] A_{\nu\,mn}+
\right. \nn\\
 & +{\bar{c}}_{nm}\left[-\square-\frac{4\xi}{a^2} \sin^2\left(\frac{\pi(n-m)}{N}\right)\right] c_{mn}+\\
& +{H}_{nm}\left[-\square-\frac{4\xi}{a^2} \sin^2\left(\frac{\pi(n-m)}{N}\right)\right] H_{mn}+ \nn\\
&\label{Squad}\left. {\bar{\psi}}_{nm}\left[
i \partialslash + i c \frac{\gamma_\ast}{a} \sin \left(\frac{2\pi (n-m)}{N}\right)- \left(m+\frac{2r c}{a} \sin^2\left[\frac{\pi (n-m)}{N}\right]\right)
\right]\psi_{mn}
 \right\rbrace\,. \nn
\end{align}
and since to the order to which we are working the interactions between the modes are being ignored, we should set $c=1$, which we do for the rest of this section.

At this stage it becomes clear how the spectrum of the theory can be independent of the coefficients in $V$ so long as the vacuum is center symmetric. Consider for instance the term 
\begin{equation*}
 \frac{c_{1}}{a^{2}}\, \mbox{Tr}( \phi) \mbox{Tr} (\phi^\dagger)
\end{equation*}
in $V$, with $c_{1} =\mathcal{O}(N_{c}^{0})$.  Expanding $\phi$ around $\Omega$, making use of the fact that $H$ is Hermitian and traceless, and switching to canonical normalization, this term makes the contribution
\begin{equation*}
 \frac{c_{1}}{N_{c}a^{2}}\, \left[ \tr ( \Omega H) \mbox{Tr} (\Omega^\dagger H)-\frac{1}{2}\tr (\Omega H^2) \tr (\Omega^\dagger)-\frac{1}{2}\tr (\Omega^\dagger H^2) \tr (\Omega)  \right]
\end{equation*}
to the quadratic action.  If it were the case that $\Omega = 1_{N_{c} \times N_{c}}$, then one would have $\tr \Omega = N_{c}$, and the second and third terms in the expression above would shift the mass of the $H$ modes at leading order.  However, if $\tr \Omega =  0$, as is the case so long as center symmetry is preserved, the second two terms above vanish, and the first term makes a contribution to the mass only at order $\mathcal{O}(N_{c}^{-1})$ relative to Equation~\eqref{eq:QuadraticAction}.  It is simple to see that the same results hold for the other terms in $V$:  as long as the theory is in a center-symmetric vacuum, the values of the coefficients in $V$ do not affect the center-symmetric observables in the theory to leading order in the $1/N_{c}$ expansion.  This is a simple illustration of the general argument to this regard given in~\cite{Unsal:2008ch}.

Coming back to our action, Equation~\eqref{eq:QuadraticAction}, we can read off the dispersion relations of the various modes in the theory. The longitudinal W-bosons ($A_\mu \sim p_\mu$) have mass
\begin{equation}
 p_0^2 = p_1^2 + \xi \frac{4}{a^2} \sin^2\left(\frac{\pi (n-m)}{N}\right)\,,
\end{equation}
and the same relation holds both for the scalars $H$ and the ghosts $c,\bar{c}$, so they all decouple in the unitary gauge $\xi\rightarrow \infty$.
The transverse W-bosons ($A_\mu \sim \epsilon_{\mu\nu} p^\nu$) are the physical ones:
\begin{equation}
\epsilon_{\rm b}(p_{1}, n,m)^{2} \equiv  p_0^2 = p_1^2 +  \frac{4}{a^2} \sin^2 \left(\frac{\pi (n-m)}{N}\right)\,,
\end{equation}
and the Majorana fermion dispersion relation gets contributions from the Wilson term and the bare mass:
\begin{equation}
 \epsilon_{\rm f}(p_{1},n m)^{2} \equiv  p_0^2 = p_1^2 + \frac{c^{2}}{a^2} \sin^2 \left(\frac{2\pi (n-m)}{N}\right) + \left[m+\frac{2r c }{a} \sin^2 \left(\frac{2\pi (n-m)}{N}\right)\right]^2\,.
\end{equation}

\subsection{Vacuum energy}
\label{sec:VacuumEnergy}
With the dispersion relations in hand, we are finally in a position to write down an expression for the vacuum energy density to leading order in an expansion in $\lambda_{3} L N_{c}$:
\begin{align}
\label{eq:VacEnergyLattice}
E_{\rm vac} &= E_{\rm b} + E_{\rm f} \\
 &=\sum^{N_{c}}_{1 \le n \le m}\int d p_{1} \sqrt{ p_1^2 +  \frac{4}{a^2} \sin^2 \left(\frac{\pi (n-m)}{N}\right)} + \\
&-\sum^{N_{c}}_{1 \le n \le m}\int d p_{1} \sqrt{ p_1^2 + \frac{c^{2}}{a^2} \sin^2 \left(\frac{2\pi (n-m)}{N}\right) + \left[m+\frac{2 r c}{a} \sin^2 \left(\frac{2\pi (n-m)}{N}\right)\right]^2} \nn
\end{align}
where the boson, $E_{b} = \sum_{p, n, m} \epsilon_{\rm b}(p, n, m)$, and fermion modes, $E_{f} = \sum_{p, n, m} \epsilon_{\rm b}(p, n, m)$,  contribute as usual with opposite signs. We note that the Higgs and longitudinal bosons cancel precisely the ghosts contribution to the vacuum energy since the ghosts come as complex scalars with fermionic statistics.   The $p_{1}$ integrals in the expressions above diverge for large $p_{1}$, and the expression is to be understood as having an implicit UV regulator about which we will have more to say shortly, while the sums over $n,m$ are finite.

Clearly, $E_{\rm vac} \neq 0$ for generic values of the parameters.  How can we reconcile this with our 3D expectation that $E_{\rm vac} = 0$ for massless fermions, on account of the $\mathcal{N}=1$ supersymmetry that emerges at $m=0$?  The key is to recall that the 2D theory is supposed to be large-$N_{c}$ equivalent to a 3D theory on a lattice in the circle direction, not directly to the continuum $\mathbb{R}^{1,1} \times S^{1}$ theory.   The lattice discretization we have chosen breaks supersymmetry, most directly thanks to the breaking of the $U(1)$ translation symmetry to its $\mathbb{Z}_{\Gamma}$ subgroup.  In an appropriate continuum limit, however, we expect the restoration of the full translation symmetry, up to corrections proportional to positive powers of the lattice spacing.  

Hence to see the fate of $E_{\rm vac}$, we should see how Equation~\eqref{eq:VacEnergyLattice} behaves in the continuum limit.  Indeed, we recall that for field theories on a spatial lattice, the dependence of the dispersion relations on the momenta is trigonometric, and the momentum integrations are restricted to the Brillouin zone.  In particular, we recall that for a free scalar field with mass $M$ on $\mathbb{R}^{1,2}$, with the $x_{2}$ direction on a lattice with spacing $a$, the dispersion relation is \cite{montvay1997quantum}
\begin{align}
p_{0}^{2} = M^{2} + p_{1}^{2} + \frac{4}{a^{2}} \sin^{2}\left(\frac{a p_{2}}{2} \right)
\end{align}
with $p_{2} \in (-\pi/a, \pi/a]$.  For small $a p_{1}$ this is just $p_{0}^{2} = M^{2} + p_{1}^{2} + p_{2}^{2}$,  but there are corrections in powers of $a$, and for large momenta ($\sim 1/a$) the the propagator is very different, as one could expect since the lattice acts as a UV regulator for $p_{2}$.

Armed with this observation, we can see the mapping of parameters necessary to interpret Equation~\eqref{eq:VacEnergyLattice} in terms of the 3D theory.  Consider first the contribution to the vacuum energy from the gauge-Higgs-ghost sector, which takes the form of 
\begin{align}
E_{\rm b} =\sum^{N_{c}}_{1 \le n \le m}\int d p_{1} \sqrt{ p_1^2 +  \frac{4}{a^2} \sin^2 \left(\frac{\pi (n-m)}{N}\right)} .
\end{align}
We now suppose that $N_{c} = \Gamma N$ for some $\Gamma, N_{c} \gg 1$, as can certainly be arranged for sufficiently large $N_{c}$, and define
\begin{align}
\frac{a p_{2}}{2} \equiv \frac{\pi (n-m)}{N_{c}}  \Rightarrow p_{2} = \frac{2\pi (n-m)}{N_{c}a} = \frac{2\pi}{L} \frac{n-m}{N} ,
\end{align}
where we recalled the relation $L = \Gamma a$ between the lattice size and the lattice spacing.  An expansion of $E_{\rm b}$ in $a p_{2} \ll 1$ now yields
\begin{align}
E_{\rm b} =\sum^{N_{c}}_{1 \le n \le m}\int d p_{1} \sqrt{ p_1^2 + p^{2}_{2}[n, m]} + \ldots.
\end{align}
and we finally see that it is the color sums in $E_{\rm b}$ which encode the third dimension.  For instance, if we consider $n,m$ such that $|n-m| \mod N = 0$, we pick up the contributions to $E_{\rm b}$ which would arise from a KK tower for a scalar on a circle of circumference $L$.  But of course the fact that the sum also includes momenta with $|n-m| \ll N$ means that $E_{\rm b}$ also includes contributions from momenta much less than the naive KK scale, so that the gap between the zero modes and the non-zero modes is actually $\frac{2\pi}{L N}$.  This is precisely the same as what happens in a perturbative calculation of the gauge-Higgs-ghost contribution to the vacuum energy the 3D theory in a center symmetric vacuum.    In both the 2D and 3D theories the KK scale in a center-symmetric vacuum becomes $\frac{2\pi}{L N}$, with the momenta becoming continuous and gapless in a large $N_{c},N$ limit with $L$ fixed, which is also the limit in which a perturbation theory becomes unreliable, since it is the volume-independent regime.  On the other hand, if one takes a non-'t Hooft large $N_{c}$ limit with $L \sim N_{c}^{-1}$, then the KK scale becomes stabilized and order $\mathcal{O}(N_{c}^{0})$, and both theories become weakly coupled.  The fact that the matching is between theories with different numbers of colors $N_{c}, N$ is expected from orbifold equivalence (and falls out naturally from volume-expanding orbifold projections), and harmless so long as $N_{c}, N \gg 1$, since the spectrum must become $N, N_{c}$-independent so long as $N, N_{c}$ are large enough and the large $N$ limit exists.

Finally, we consider $E_{\rm f}$, and use the same parameter identifications as in the boson case to see the connection to the 3D theory.  As is usually the case, the fermion case is more subtle.  First, to the order to which we are working, we must take $c=1$, and will do so for the rest of this section.  If we also take $m = r = 0$, then the extra factor of two in the the argument of the $\sin^{2}$ in $E_{\rm f}$ together with the limits on the color sum implies that the near-continuum expansion of the $\sin^{2}$ includes contributions both from near $|a p_{2}| \approx 0$ and $|a p_{2}| \approx \pi$.   Hence
\begin{align}
E_{\rm f}(m=0, r=0) =-2\sum^{N_{c}}_{1 \le n \le m}\int d p_{1} \sqrt{ p_1^2 + p^{2}_{2}[n, m]} + \ldots.
\end{align}
and $E_{b} + E_{f} \neq 0$.  Indeed, the vacuum energy turns out to be $E_{b} + E_{f} = - E_{f}$, and it is as if there were two flavors of Majorana fermions in the theory, rather than one. In fact, this is exactly right.  If $r=0$, the 3D lattice theory to which the 2D theory is large-$N_{c}$ equivalent suffers from fermion doubling, and for $m=0$ there are two physical `flavors' of fermions rather than the one flavor one thought one was discretizing.  

If we turn on $r > 0$, things change dramatically.  The expansion of the trigonometric factors in $E_{\rm f}$ near $a p_{2} \approx \pi$  picks up an order $1/a^{2}$ mass-squared term proportional to $r$, while at the same time the Wilson term makes no contribution to the mass of the modes with $a p_{2} \approx 0$ to this order in perturbation theory.  Hence the doubler mode becomes much heavier than the `continuum' mode we wanted, and so long as the momentum cutoffs implicit in $E_{\rm vac}$ are lower than $1/a$,  the contribution of the doubler mode to the continuum physics can be neglected.  Of course, this was precisely the motivation for including the Wilson term in the lattice theory in the first place:  Wilson terms prevent fermion doubling.

Hence we see that to the order to which we are working and with a sensible choice of UV cutoff,  we find
\begin{align}
E_{\rm vac}(m=0, r\neq0) &= 0 ,
\end{align}
as one would expect for a theory which has emergent supersymmetry in the continuum limit for a massless Majorana fermion.  If we were to go to the next order in perturbation theory, we expect that for $r \neq 0$, to get the physical fermion mass to be zero the bare fermion mass $m$ would have to be tuned appropriately, since the Wilson term breaks the same symmetries as the mass term and hence induces additive mass renormalization \cite{montvay1997quantum}.  At the next order in perturbation theory we would also have to adjust the value of $c$ away from $c=1$. These observations serve to re-emphasize the points we have made before, which are that in general $m \neq M, c \neq 1$.

\section{Bosonization of adjoint fermions}
\label{sec:AdjointBosonization}
To bosonize the theory described by Equation~\eqref{eq:2DTheory}, we need to know the bosonization `dictionary' for massive adjoint fermions coupled $SU(N_{c})$ gauge fields and scalar fields.  A bosonization recipe for fermions with non-Abelian color or flavor symmetries was invented in~\cite{Witten:1983ar}, and rederived using a constructive path integral framework in \cite{DiVecchia:1984df,DiVecchia:1984mf,Gonzales:1984zw,Burgess:1994ef}.  These path integral derivations of the bosonization dictionary are based on the beautiful explicit calculation of the functional determinant of fermions coupled to gauge fields of Polyakov and Wiegmann \cite{Polyakov:1983tt,Polyakov:1984et}.  In Section~\ref{sec:BosonizationKineticTerms}, we follow \cite{Polyakov:1983tt,Polyakov:1984et} to compute the functional determinant for adjoint fermions coupled to $SU(N_{c})$ gauge fields.  Once this is obtained, the methods of \cite{DiVecchia:1984df,DiVecchia:1984mf,Gonzales:1984zw,Burgess:1994ef} yield the bosonized representation of adjoint fermions coupled to gauge fields without any mass or Yukawa terms.  In Section~\ref{sec:MassYukawaGeneral} we explain how to take into account mass and Yukawa terms, which is necessary to enable the bosonization of the theory described in Section~\ref{sec:DimensionalReduction}.

\subsection{Evaluation of massless fermion action}
\label{sec:BosonizationKineticTerms}
We first suppress the mass terms and scalar couplings, and consider free Majorana fermions $\psi$ in the adjoint representation of $SU(N_{c})$, described by the Lagrangian
\begin{align}
\label{eq:AdjFermionLagrangian}
\mathcal{L}_{\rm adj} = \tr[ \bar{\psi} i \gamma^{\alpha}{D_{\alpha}} \psi].
 \end{align}
The $SU(N_{c})$ gauge fields $A^{\rm SU}_{\mu} = A^{a}_{\mu}t_{a}, a = 1, \ldots, N_{c}^{2}-1$  in this expression are to be thought of as background source fields with local $SU(N_c)$ transformation, and the path integral over the fermions defines the generating functional
\begin{align}
\label{eq:AdjFermionPartitionFunction}
Z[A^{\rm SU}_{\mu}] &= \int  \mathcal{D}\psi \, e^{i \int d^{2}x\, \mathcal{L}_{\rm adj}} \\
&= \det [C  (i\gamma^{\alpha} \partial_{\alpha} + A^{\rm SU}_{\mu})]^{1/2} = \exp{i \left[\tr \log W(A^{\rm SU}_{\mu})\right]}
\end{align}   
where the square root appears because we are dealing with Majorana fermions\footnote{Strictly speaking the result of integrating over Majorana fermions is the Pfaffian, which can differs from the square root of the determinant by a non-trivial sign, but this subtlety will not be important for us.}.  Our aim is to find an explicit expression for $Z[A^{\rm SU}_{\mu}]$ as the exponential of some bosonic local action.

Before discussing the derivation of the bosonic representation of $Z[A^{\rm SU}_{\mu}]$ we must discuss an important subtlety.  The issue is that  $Z[A^{\rm SU}_{\mu}]$ does not describe all of the correlation functions of currents that could be defined for Equation~\eqref{eq:AdjFermionLagrangian}.  The reason is that in the absence of $SU(N_{c})$ gauge fields,  Equation~\ref{eq:AdjFermionLagrangian} actually has an $SO(N^{2}-1)$ global symmetry under which fermions transform in the vector representation, and the theory has a large number of conserved currents that do not couple to the $A_{\mu}$ defined above, which only parametrizes an $SU(N_{c})$ subgroup of $SO(N_{c}^{2}-1)$.    Hence one could try to bosonize adjoint fermions by finding a bosonic representation of $Z[A^{\rm SO}_{\mu}]$, where $A^{\rm SO}_{\mu}$ are background $SO(N_{c}^{2}-1)$ gauge fields.  To obtain the $Z[A^{\rm SU}_{\mu}]$ generating functional above, one would then have to set to zero all of the components of $A^{\rm SO}_{\mu}$ that do not live in some chosen $SU(N_{c})$ subgroup of $SO(N_{c}^{2}-1)$.

However, given our goal of describing a theory with a bona-fide $SU(N_{c})$ gauge symmetry, for which we only actually need $Z[A^{\rm SU}_{\mu}]$, the $SO(N_{c}^{2}-1)$-symmetric approach would be overkill.  Instead, we will pursue the much more efficient approach of directly determining a bosonized description of $Z[A^{\rm SU}_{\mu}]$, so that we only explicitly work with a bosonized description that describes the degrees of freedom that play a role in the $SU(N_{c})$ gauge theory of interest.  The result of this approach will be essentially the same as the one reached in~\cite{Kutasov:1994xq}. 

These issues can be understood more explicitly in a much easier setup. If we consider a theory of just three free Majorana fermions then at the free level we would have an $SO(3)_L \times SO(3)_R$ symmetry, and the bosonization of the theory would give rise  to an $SO(3)$ WZW model \cite{Witten:1983ar}. This amounts to working out an expression for the fermion determinant coupled to an $SO(3)$ gauge field, and then using the techniques of \cite{DiVecchia:1984df,DiVecchia:1984mf,Gonzales:1984zw,Burgess:1994ef} to get the bosonized theory.  However, suppose we are actually interested in gauging an $SO(2)_V \cong U(1)$ subgroup of the  fermionic theory.   On the fermionic side, this is simply a theory of a massless Dirac fermion charged under a $U(1)$ gauge field and a neutral Majorana fermion; a gauge-invariant mass term could of course be added.  In particular since the neutral Majorana fermion does not interact with the Dirac fermion or with the gauge field, correlation functions involving both the Dirac and Majorana fermion factorize in a trivial way.  If one wanted to bosonize such a fermionic theory, one could simply use Abelian bosonization to capture the correlation functions of the Dirac fermion, and the result is a theory of a scalar field coupled to a Sine-Gordon potential and a gauge field.  This corresponds to working out an explicit expression for the fermion determinant coupled to $U(1)$ gauge field, and then applying the techniques of \cite{DiVecchia:1984df,DiVecchia:1984mf,Gonzales:1984zw,Burgess:1994ef} to get the bosonized theory.  Of course, the same physics is contained in the $SO(3)$ WZW model coupled to a $U(1)$ gauge field, but the encoding is rather complicated since the $SO(3)$ boson representation describes the full set of 
$SO(3)_L \times SO(3)_R$ correlation functions.  But so long as one is \emph{only} interested in the non-trivial gauge-invariant correlation functions involving the Dirac fermion field, the Abelian bosonization is equivalent to the full non-Abelian bosonization, with the virtue of being far simpler.

In what follows, then, we seek to obtain an expression for $Z(A^{\rm SU}_{\mu})$ directly, following the methods of Polyakov and Wiegmann. The method of \cite{Polyakov:1983tt} is based on the construction of solutions to the Ward identities of the theory in terms of bosonic variables.  To begin, note that the classical action has the vector and axial flavor symmetry $SU(N)_{V} \times SU(N)_{A}$, with $\psi$ transforming in the adjoint representation of both symmetry groups.  The vector current is by definition sourced by $A_{\mu}$:
\begin{align}
\label{eq:VariationalDerivative}
J^{V}_{\mu} = \frac{\delta W }{\delta A_{\mu}}.
\end{align}
Crucially for what follows, the two-dimensional identity $\gamma^{5} \gamma^{\mu} = \epsilon^{\mu \nu} \gamma_{\nu}$ implies that the axial current is $J^{A}_{\mu} = \epsilon_{\mu \nu} J^{V,\nu}$, which allows us to rewrite relations for $J^{A}$ in terms of $J^{V}$.  As is well-known, $J^{V}_{\mu}$ is a conserved current, while the axial current is anomalous, and the currents obey the equations
\begin{align}
\label{eq:AnomalyEquationSO}
\partial^{\mu} J^{V}_{\mu} + i\,[A^{\mu}, J^{V}_{\mu}] &= 0   \\
\partial^{\mu} J^{A}_{\mu} + i\,[A^{\mu}, J^{A}_{ \mu}] = \epsilon^{\mu \nu}\left(\partial_{\nu} J^{V}_{\mu} +i\, [A_{\nu}, J^{V}_{ \mu}] \right)&= \frac{N_{c}}{2\pi} \epsilon^{\mu \nu} F_{\mu \nu}   
\end{align}
We now follow the method of Polyakov and Wiegmann \cite{Polyakov:1983tt} to find an expression for $W(A_{\mu})$.

The approach is technically simplest in lightcone coordinates $x_{\pm} = x_{0} \pm x_{1}$, to which we now shift.
We can rewrite Equation \eqref{eq:AdjFermionLagrangian} in a $SU(N_c)_L\times SU(N_c)_R$ manifestly invariant way:
\begin{equation}
 \label{eq:AdjFermionLagrangianLR}
\mathcal{L}_{\rm adj} = \tr[ \psi_L^\ast \left(i\partial_+ - [A_+,\,.]\right) \psi_L+
 \psi_R^\ast \left(i\partial_- - [A_-,\,.]\right) \psi_R]\,.
\end{equation}
Clearly the vectorial and axial $SU(N_c)$ gauge fields can be recovered by suitable combination of left and right $SU(N_c)$ gauge fields. We now rewrite $A_{\pm}$, which live in the lie algebra of $SU(N_{c})$, in terms of group-valued variables:
\begin{align}
\label{eq:GaugeFieldAnsatz}
A_{+}= -i\,g^{-1} \partial_{+} g, \qquad A_{-}= -i\, h^{-1} \partial_{-} h
\end{align}
and $g, h$ live in the group $SU(N_{c})$, and pure gauge configurations correspond to the choice $g=h$. In fact, any elements of $SU(N_{c})$ differing by multiplication by an element of the center $G_{c}$ subgroup of $SU(N_{c})$ will lead to the same $A_{\pm}$, so the correct statement is that $g, h \in SU(N_{c})/G_{c}$.  The trick is now to find expressions for $J_{\pm}$ in terms of $g, h$ which solve Equation~\eqref{eq:AnomalyEquationSO}.  Indeed, it turns out that the desired expressions are~\cite{Polyakov:1983tt}
\begin{align}
J_{+} &= \frac{iN_{c}}{2\pi} (g^{-1} \partial_{+} g - h^{-1} \partial_{-} h) \\
 J_{-} &= \frac{iN_{c}}{2\pi} (h^{-1} \partial_{+} h - g^{-1} \partial_{-} g)
\end{align}
 It is easiest to see these expressions satisfy Equation~\eqref{eq:AnomalyEquationSO} in the axial gauge $A_{-} = 0$,  which can be accomplished by a gauge transformation that maps $g \to g h^{\dagger} = g'$. Dropping the prime, one can easily check that $J_{\pm}$ do the job by using the identity
\begin{align}
D_{+} (g^{-1} \partial_{+} g) - \partial_{-}(g^{-1} \partial_{+} g) = 0
\end{align}
where $D_{+}$ is the adjoint covariant derivative.  

With these expressions in hand, we can rewrite Equation~\eqref{eq:VariationalDerivative} as
\begin{align}
\label{eq:VariationEquation}
\delta W(g) = \int d^{2}x\, \left[ J_{-} \delta A_{+} \right] = \frac{N_{c}}{2\pi} \int d^{2}x \partial_{-} (g^{-1} \partial_{+} g) \delta g g^{-1}  .
\end{align}
The remaining step is to find the action $W(g)$ whose variation is Equation~\eqref{eq:VariationEquation}. As it happens, there is no manifestly $SU(N_{c})_{V}$-invariant 2D action for $g$ which reproduces the expected $SU(N_c)_A$ variation (i.e. the anomaly) which obeys Equation~\eqref{eq:VariationEquation}.  However, it is possible to write an action with manifest $SU(N_{c})_{V}$ invariance and desired variation properties using the trick of allowing terms in the action to be written as integrals over a three-dimensional manifold which has the 2D spacetime as a boundary.   The desired action is the Wess-Zumino-Witten action \cite{Witten:1983ar,Polyakov:1983tt}
\begin{align}
W(g) = \frac{N_{c}}{8 \pi}\left[ \int d^{2}x \tr(\partial_{\mu} g^{-1} \partial_{\mu} g) +\frac{2}{3}\int_{B}d^{2}x\, d\xi \,\epsilon^{ABC} \tr ( g^{-1} \partial_{A} g g^{-1} \partial_{B} g g^{-1} \partial_{C} g) \right]
\end{align}
where the integral in the third term runs over $D = \mathbb{R}^{1,1} \times [0,1]$, which in Euclidean space with $\mathbb{R}^{2} \to S^{2}$ would simply be a solid ball $S^{2} \times [0,1]$.  The extension of the map $g: S^{2} \to SU(N_{c})/G_{c}$ to $S^{2} \times [0,1] \to SU(N_{c})/G_{c}$ does not have any topological obstructions since $\pi_{2}(SU(N_{c})/G_{c}) = 0$, and the topologically distinct extensions to $D$, classified by $\pi_{3}(SU(N_{c})/G_{c}) = \mathbb{Z}$, all make the same contribution to the path integral provided $N_{c} \in \mathbb{Z}$.   It is also straightforward to switch back to a general gauge as explained in~\cite{Polyakov:1984et}, getting a result that depends on $g$ and $h$ and turns out to be writable as $W(g h^{\dag})$.  For what follows we assume the switch to the general gauge has been made, rename $g h^{\dag} = g'$ and drop the prime.

Finally, having obtained an expression for $W(g)$, we can insert it into the path-integral bosonization/dualization machinery of \cite{DiVecchia:1984df,DiVecchia:1984mf,Gonzales:1984zw,Burgess:1994ef}, which re-expresses $Z(A)$ as a functional integral over $g$, so that
\begin{align}
Z(A) =&\notag \int dg \exp\left(\frac{i N_{c}}{8 \pi} \int d^{2}x \tr(D_{\mu} g^{-1} D^{\mu} g) +i\,N_c \tilde{\Gamma} (g,A)\right)\,,\\
\tilde{\Gamma}(g,A)=&\label{eq:MasslessBosonization}\frac{1}{12\pi}\int_{B}d^{2}x\, d\xi \,\epsilon^{ABC} \tr ( g^{-1} \partial_{A} g g^{-1} \partial_{B} g g^{-1} \partial_{C} g)+\\
&\notag-\frac{1}{4\pi}\int d^2x\,\epsilon^{\mu\nu}\tr [i A_\mu (g^{-1}\partial_\nu g-g\,\partial_\nu g^{-1}+i g^{-1}A_\nu g)] 
\end{align}
where $D_{\mu} = \partial_{\mu} + i [A_{\mu}, \cdot]$, and $g\in SU(N_{c})/\mathbb{Z}_{N_{c}}$.  This is the desired bosonized expression for adjoint fermions coupled to $SU(N_{c})$ gauge fields \cite{Frishman:2010zz}.

By construction, the bosonized theory transforms in the same way as the fermionic theory under local $SU(N_{c})_{L} \times SU(N_{c})_{R}$ transformations, which act on $g$ as $g \to L g R^{-1}$.  Just like in the fermionic description, only the vectorial transformations (i.e. $L=R$) can be gauged. Of course, to see the non-invariance of $W$ under local axial transformations requires a one-loop calculation in the fermionic language, while in the bosonic description this can be seen at the classical level.   This is one of the key features of bosonization:  one-loop physics in the fermionic language turn into classical physics in the bosonized description.  

Finally, we note that the fact that $g$ lives in $SU(N_{c})/G_{c}$ rather than $SU(N_{c})$ is important in ensuring that the bosonized action does not have a non-trivial symmetry under multiplication of $g$ by elements of $G_{c}$, which is fortunate since there is no such symmetry in the fermionic theory.

\begin{table}[t]
\centering
\begin{tabular}{ c |  c  c  c c }
\toprule
\multicolumn{5}{c}{Transformation Properties} \\
\hline
Symmetry & $\psi_{L}$ & $\psi_{R}$ & $\phi$ & $g$ \\
\hline
$SU(N_{c})_{V}$ & $V\psi_{L}V^{\dag}$ & $V\psi_{R}V^{\dag}$ & $V\phi
V^{\dag}$ & $V g V^{\dag}$ \\
$SU(N_{c})_{A}$ & $A\psi_{L}A^{\dag}$ & $A^\dag \psi_{R}A$ & $\phi$ & $A g
A$ \\
\hline
$SU(N_{c})_{L}$ & $L\psi_{L}L^{\dag}$ & $\psi_{R}$ & $\phi$ & $L g$ \\
$SU(N_{c})_{R}$ & $\psi_{L}$ & $R\psi_{R}R^{\dag}$ & $\phi$ & $g R^{\dag}$
\\
\hline
$\mathbb{Z}_{N_{c}}$ & $\psi_{L}$ & $\psi_{R}$ & $\omega \phi$ & $g
$ \\
\hline
\end{tabular}
\caption{Transformation properties of matter fields in the bosonic and
fermionic 2D theories.}
\label{t:symm}
\end{table}

\subsection{Mass and Yukawa terms}
\label{sec:MassYukawaGeneral}
To bosonize the theory described by Equation~\eqref{eq:2DTheory}, we need to find a bosonization `dictionary' for massive adjoint fermions coupled $SU(N_{c})$ gauge fields and scalar fields.  Hence we now discuss the bosonization of mass and Yukawa-like terms in fermionic theories.

The desired entries in the bosonization dictionary are most easily determined using spurion analysis\footnote{For an interesting alternative prescription for the bosonization of mass terms of adjoint fermions see \cite{Smilga:1996dn}.  The relation between our results and those of \cite{Smilga:1996dn} is unclear to us.}.  The most general Yukawa-like term which can be made consistent with 
\begin{align*}
 \psi_L &\rightarrow L \psi_L L^\dagger\,,\\
\psi_R &\rightarrow R \psi_R R^\dagger\,,
\end{align*}
 with $L,R\in SU(N_{c})_{L} \times SU(N_{c})_{R}$ [the full symmetry group of 
Equation~\eqref{eq:AdjFermionLagrangianLR}] can be written as
\begin{align}
\mathcal{L}_{F,y} = \tr \bar{\psi} A \psi B + h.c. = \tr \bar{\psi}_{L} A \psi_{R} B + h.c.
\end{align}
where $A, B$ are spurion fields transforming as $A \to L A R^\dagger, B \to R B L^\dagger$.  There is also a symmetry $A \to z A, B \to z^{-1}B, z \in \mathbb{C} \setminus \{0\}$ which acts only on the spurion fields.   Note that such a term is a relevant deformation of the fermionic action, since the mass dimension of $A, B$ must be such that $[A]+[B] = 1$, and for particular choices of $A, B$ one recovers the mass terms and $\phi$-coupling terms in Equation~\eqref{eq:2DTheory}.  

In the bosonized description of the theory the dynamical field is $g$, which transforms under $SU(N_{c})_{L} \times SU(N_{c})_{R}$ as $g \to L g R^{\dag}$.  Since $\mathcal{L}_{F,y}$ is a relevant deformation of the fermionic theory, the same must be true for the term(s) induced by turning on the spurion fields in the bosonized theory.  As it happens, the only relevant term involving $g, A$, and $B$ which is allowed by the symmetries of $g$ and the spurion fields is
\begin{align}
\mathcal{L}_{B,y} = \mu \left[ \tr(g B)\tr(g^{\dag} A) + h.c.\right]
\end{align}
where $\mu$ is an undetermined positive parameter with mass dimension $[\mu]=1$.  The sign of $\mu$ is determined by the demanding that if $A,B$ are set to constant values, as would be the case if one was bosonizing a mass term, so that $A=B= \sqrt{m}, m>0$, the bosonized theory must be stable, with a non-tachyonic mass term for fluctuations of $g$ around the identity.   

The other relevant term consistent with the $SU(N_{c})_{L} \times SU(N_{c})_{R}$ symmetry, which is
\begin{align}
\mathcal{L}'_{B,y} = \mu' \left[ \tr(g A g B^{\dag})+ h.c.\right], 
\end{align}
is ruled out by the $U(1)$ symmetry acting on $A,B$. 
Finally, we note that the fact that the bosonized action is determined only up to a dimensionful parameter reflects the fact that the 
fermionic path integral with mass or Yukawa terms turned on is only really well-defined once a particular regularization scheme is chosen, leading to the fact e.g. $\langle \bar{\psi} \psi \rangle$ is scheme-dependent, and is a standard feature of bosonization both in Abelian and non-Abelian cases.  The fact that this issue shows up already at the level of the classical action in the bosonized theory is another reflection of the fact that one-loop physics in the fermionic language becomes classical physics in the bosonic theory.

\begin{table}[t]
\centering
\begin{tabular}{ c | c  c }
\toprule
\multicolumn{3}{c}{Bosonization Dictionary} \\
\hline
$J^{a}_{+}$ &$\tr\, \bar{\psi} \gamma_{+} t^{a} \psi$ & $\frac{i N_{c}}{2
\pi} \tr\, g^{\dag} \partial_{+}g\, t^{a}$  \\
$J^{a}_{-}$ &$\tr \,\bar{\psi} \gamma_{-} t^{a} \psi$ & $\frac{i N_{c}}{2
\pi} \tr \,\partial_-g^\dag g \,t^{a}$ \\
$\mathcal{O}_{\rm Yukawa}$ &$\tr\, \bar{\psi}A \psi B + \mathrm{h.c.}$ &
 $\mu \left[ \tr(g B)\tr(g^{\dag} A) + h.c.\right]$  \\
\hline
\end{tabular}
\caption{Bosonization dictionary for $SU(N_{c})$ adjoint fermions with
Yukawa and mass terms.}
\label{table:BosonizationDictionary}
\end{table}

\section{From 3D gauge theory with fermions to a bosonized 2D theory}
\label{sec:3DBosonization}
With the results assembled in Section~\ref{sec:DimensionalReduction} and Section~\ref{sec:AdjointBosonization}, we finally have all of the ingredients we need to bosonize 3D $SU(N_{c})$ gauge theory with adjoint fermions at large $N_{c}$ using the dictionary summarized in Table \ref{table:BosonizationDictionary}.  Aside from the mass and Yukawa terms coupling $\psi$ with $\phi$, the action of the bosonized description of Equation~\ref{eq:2DTheory} will be given by Equation~\eqref{eq:MasslessBosonization}.   Our task in Section~\ref{sec:CouplingGandPhi} is to derive the terms in the bosonized action coupling $g$ with $\phi$ which aredue to the couplings between $\psi$ and $\phi$ in the 2D theory.  Finally, in Section~\ref{sec:BosonizationFinalForm} we put everything together to exhibit the full bosonized action.

\subsection{Couplings of $\phi$ and $g$}
\label{sec:CouplingGandPhi}
To make use of the analysis carried out in the previous section we need to rewrite the fermionic interactions in the spurion formalism. For the once-and-future kinetic term, we have
\begin{equation*}
 \frac{-c}{2a} \mbox{Tr} \left([\bar{\psi}, \phi] \gamma_5 \{ \psi, \phi^{\dagger}\}\right)
\end{equation*}
Anticommuting the fermions, using the unitarity condition: $\phi\phi^\dagger=1$ and the cyclic properties of traces, we obtain the form:
\begin{equation}
 \frac{c}{a}\mbox{Tr}\left( {\bar{\psi}}_L \phi^\dagger \psi_R\,\phi- {\bar{\psi}}_L \phi\, \psi_R \phi^\dagger\right)\,,
\end{equation}
and these two terms are separately Hermitian (recall that we are dealing with Majorana fermions).
We can repeat the same procedure for the Wilson and mass term:
\begin{equation*}
   \mbox{Tr}\left(\frac{r}{2a} [\bar{\psi}, \phi] [ \psi, \phi^{\dagger}] -m\bar{\psi}\psi\right)
\end{equation*}
and using the same properties as above we get
\begin{equation}
\label{eq:AllFermions}
 \mbox{Tr} \left(\frac{r c}{a}  {\bar{\psi}}_L \phi^\dagger\psi_R\,\phi+ \frac{r c}{a}  {\bar{\psi}}_L \phi \,\psi_R \phi^\dagger- 2\left(\frac{r c}{a} + m\right)\bar{\psi}_L \psi_R\right).
\end{equation}

At this point we have to identify the spurion fields allowed by the transformation properties given in Table \ref{t:symm}. In the generic case with both $m,r \neq 0$ we can introduce two spurions: 
one for the combined mass term containing the bare mass and Wilson mass and a second for the once and future kinetic term together with the Wilson term\footnote{Note that as is explained in e.g.~\cite{montvay1997quantum}, the Wilson term coefficient should be chosen to obey $|r| \leq 1$ to ensure reflection positivity.}
\begin{align}
\mbox{Tr} \left({\bar{\psi}}_L A_1 \psi_R B_1 + {\bar{\psi}}_L A_2 \psi_R B_2 +h.c.\right)
\end{align}
with
\begin{align}
\begin{array}{l l}
A_1= \frac{i}{\sqrt{a}} \sqrt{am+r c}\quad   & B_1 =\frac{i}{\sqrt{a}} \sqrt{am+r c} \\
A_2= \sqrt{\frac{c}{2a}} \left(\phi\sqrt{1-r} +\phi^{\dag}\sqrt{1+r} \right)\quad  &B_2 =\sqrt{\frac{c}{2a}}\left(\phi \sqrt{1+r}- \phi^{\dag}\sqrt{1-r}
\right)
\end{array}
\end{align}
The fact that we need only two spurion fields, and not {\it e.g.} three, can also be seen by using the symmetries of the original fermionic theory such as $\psi_{L} \to - \psi_{L}, \phi \to \phi^{\dag}, m\to-m, r\to-r$.
 
Using the machinery built up in the previous section we arrive at the bosonized counterpart of the desired terms:
\begin{align}
\label{eq:YukawaRGeneral}
 \delta \mathcal{L}= (m+\frac{r}{a})\,\mu_{1} \,\mbox{Tr} \,g\, \mbox{Tr}\, g^\dagger -\frac{\mu_{2}  c (r-1)}{2a} \mbox{Tr} (g^\dagger \phi) \mbox{Tr} (g \phi^\dagger)-
\frac{\mu_{2} c(r+1)}{2a} \mbox{Tr} (g \phi) \mbox{Tr} (g^\dagger \phi^\dagger)\,.
\end{align}
Had we started with the standard value of $r$ in lattice calculations, $r=1$, then we could have used just one spurion consistently with all the transformation of Table \ref{t:symm}:
\begin{align}
\mbox{Tr} \left({\bar{\psi}}_L A \psi_R B +h.c.\right)
\end{align}
with
\begin{align}
& A= \sqrt{\frac{c}{a}} \left(\phi^\dagger+i\sqrt{\frac{am}{c}+1}\right), \quad B =\sqrt{\frac{c}{a}} \left(\phi+i\sqrt{\frac{am}{c}+1}\right) \,,
\end{align}
yielding a particularly simple contribution to the bosonized Lagrangian:
\begin{align}
\label{eq:YukawaR1}
 \delta \mathcal{L}= \left(\frac{ma+c}{a}\right) \mu \,\mbox{Tr} \,g\, \mbox{Tr}\, g^\dagger-
\frac{\mu c}{a} \mbox{Tr} (g \phi) \mbox{Tr} (g^\dagger \phi^\dagger) \,.
\end{align}
Finally, we note that $\mu, \mu_{1}, \mu_{2}$ must scale as $N_{c}^{0}$ in the large $N_{c}$ limit,  due to the presence of two color traces in Equations~\eqref{eq:YukawaRGeneral} and \eqref{eq:YukawaR1},  so that the contributions of these terms to the bosonized action have the same large $N_{c}$ scaling as the corresponding terms in the fermionic action.  

\subsection{Bosonized form of 3D YM with an adjoint fermion}
\label{sec:BosonizationFinalForm}

We now have all of the ingredients we need to bosonize the theory described by Eq.~\eqref{eq:2DTheory}.  Making the simplifying choice $r=1$, and combining the results above with those of Section~\ref{sec:BosonizationKineticTerms}, the bosonized version of $S_{2D}$:
\begin{align}
\label{eq:BosonizedAction}
 S^{b}_{2D}= & \int d^2x \left\lbrace \,\tr \left(\frac{-1}{2g_2^2} F^2+
\frac{1}{a^2g_2^2} \vert D_\mu \phi\vert^2 + \frac{N}{8\pi} \vert D_\mu g\vert^2
\right) \right. \nn  \\
&\left. + \tilde{m}^{2}  \,\tr \,g\, \tr \, g^\dagger - \frac{\tilde{c}}{a^{2}} \tr  (g \phi) \tr (g^\dagger \phi^\dagger)
\right\rbrace+N\tilde{\Gamma}(g,A)
\end{align}
where we have defined $\tilde{m}^{2} \equiv \mu(ma+c)a^{-1}$ and $\tilde{c} \equiv c \mu a$.  Together with the dimensional reduction formula in Eq.~\eqref{eq:2DTheory}, this is our main result.  The fact that bosonization is an exact procedure in two dimensions together with the large $N_{c}$ equivalence of Eq.~\eqref{eq:2DTheory} to the original 3D gauge theory implies that large $N_{c}$ 3D gauge theories coupled to adjoint fermions can be bosonized, with the resulting gauge theories written in terms of bosonic fields living in two dimensions.

As the notation we chose above is meant to suggest, $\tilde{m}$ is a mass parameter which is related to the original bare fermion mass $M$, while $\tilde{c}$ plays the same role as the `speed of light' tuning parameter $c$ in the dimensionally reduced theory.  Indeed, as we emphasized in Section~\ref{sec:Orbifolding}, the values of $m, c$ (and hence $\tilde{m}, \tilde{c}$) which correspond to a Lorentz-invariant continuum limit with a given physical quark mass must in general be determined non-perturbartively when the theory is strongly coupled, and get corrections order by order in perturbation theory when the theory is weakly coupled.  This perspective makes the appearance of the parameter $\mu$ in the relations between $\tilde{m}, \tilde{c}$ and $m, c$ especially natural,  since the bosonized action encodes physics which is arises at the  one-loop level in the original fermionic action. 

For the sake of completeness we also present the generic case with $m \neq 0, r \neq0$:
\begin{align}
 S= &\int d^2x \left\lbrace \,\tr \left(\frac{-1}{2g_2^2} F^2+
\frac{1}{a^2g_2^2} \vert D_\mu \phi\vert^2 + \frac{N}{8\pi} \vert D_\mu g\vert^2
\right) \right.   \\
&\notag\left. + \tilde{m}'^{2}  \,\tr \,g\, \tr \, g^\dagger 
- \frac{\tilde{c}'(r-1)}{2a^{2}} \tr(g^\dagger \phi) \tr (g \phi^\dagger) -
\frac{\tilde{c}' (r+1)}{2a^{2}}\tr (g \phi) \tr(g^\dagger \phi^\dagger)
\right\rbrace+N\tilde{\Gamma}(g,A)
\end{align}
where $\tilde{m}'^{2} \equiv \mu_{1}(ma+c r)a^{-1}$ and $\tilde{c}' \equiv c \mu_{2} a$, with the same comments concerning the connection of $\tilde{m}', \tilde{c}'$ to $m, c$ as above applying.

\section{Conclusions}
\label{sec:Conclusions}
The recent realization that D-dimensional $SU(N_{c})$ gauge theories compactified on spatial tori {\it e.g.} $\mathbb{R}^{1,1}\times S^{D-2}$  with adjoint fermions stay in a center-symmetric phase for arbitrary torus size $L$ has important consequences for both analytic and numerical studies of gauge theories.  Analytic developments have so far mainly focused on the $LN\Lambda_{\rm strong} \ll 1$ regime, which is center-symmetric but not volume-independent, and have very recently lead to dramatic progress toward a non-perturbative continuum definition of a class of gauge theories\cite{Argyres:2012ka,Argyres:2012vv}.  The main applications of the notion of large $N$ volume independence, which holds when $L N \Lambda_{\rm strong} \gtrsim 1$, has so far been for Monte Carlo simulations, since it means Eguchi-Kawai reduction to one plaquette is valid for such theories \cite{Hietanen:2009ex,Bringoltz:2009kb,Bringoltz:2009by,Azeyanagi:2010ne,Catterall:2010gx,Hietanen:2010bz,Hietanen:2010fx,Bringoltz:2011by,Hietanen:2012ma}.  Our results here suggest that the notions of large $N_{c}$ volume independence and large $N_{c}$ dimensional reduction may also be quite useful for obtaining novel analytical results.

Our main results are Equation~\eqref{eq:2DTheory} and Equation~\eqref{eq:BosonizedAction}.  The notion of large $N_{c}$ dimensional reduction has appeared before in \cite{Bedaque:2009md,Poppitz:2009fm} in the context of reduction from a 4D theory to a 3D one, but we believe the presentation here clarifies a number of aspects of large $N_{c}$ dimensional reduction.  As an application of this technology, we showed that (perhaps unexpectedly) 3D $SU(N_{c})$ gauge theory with one flavor of adjoint fermions can be bosonized, with the bosonized action given in Equation~\eqref{eq:BosonizedAction}.  A remarkable feature of this result is that the bosonized action corresponding to the large $N$ three-dimensional theory is local, but lives in two dimensions.

There are a large number of possible applications and extensions of our results.  For instance, while we focused our discussion on 3D $SU(N_{c})$ gauge theory with $N_{f}=1$ adjoint fermions, the techniques we discussed are clearly more general.  For instance, one could bosonize 4D gauge theories by working on $\mathbb{R}^{1,1} \times S^{1} \times S^{1}$, discretizing on a lattice with $\Gamma^{2}$ sites, performing the large $N_{c}$ dimensional reduction, and then applying a generalization  of the adjoint fermion bosonization dictionary to multiple flavors to find the 2D bosonic actions corresponding to 4D  large $N_{c}$ gauge theories.  It would be very interesting to work out the explicit form of such bosonized theories.

A crucial issue which deserves further studies is the analysis of the spectrum beyond the quadratic level.
Clearly this is a difficult task but in principle we could combine the expansion around the center symmetric vacuum with the powerful tool of discretized light-cone quantization (DLCQ) to analyse the full interacting spectrum. The idea is to compactify also the light cone space $x^-$ so that all the momenta will be quantized and the 2D Hamiltonian could be diagonalized numerically at large-$N$ using only single trace states. Previous studies  (see {\it e.g.}~\cite{Demeterfi:1993rs,Bhanot:1993xp,Matsumura:1995kw,Harada:2004ck,Dorigoni:2010jv}) were always carried out in the trivial vacuum, while in our analysis of the dispersion relations in the free case, we saw that the presence of a non-trivial Polyakov loop turns out to be crucial to encode the momenta in the reduced direction.  Hence it would be especially interesting to repeat the DLCQ calculations using the center-symmetric vacuum.

Another interesting direction is to start with D-dimensional gauge theories with extended supersymmetry, and attempt to define lattice regularizations on $T^{D-2}$ which preserve some supersymmetry at finite lattice spacing (see \cite{Catterall:2009it} for a review of lattice supersymmetry).  This should result in a version of large $N_{c}$ dimensional reduction with manifest supersymmetry and might enable some interesting calculations which would be analytically intractable using the simple-minded regularization we adopted here.  It should also be very interesting take advantage of the super-renormalizability of 3D gauge theories to do the perturbation theory calculations necessary to find the expressions for $m, c$ which lead to the desired continuum limit \cite{Reisz:1987da,Golterman:1988ta,Catterall:2000rv,Kaplan:2003uh,Giedt:2004vb,Elliott:2005bd,Elliott:2007bt}.

As a final example of an application of our results, one could try to extract some phenomenology from the bosonized form of the 3D theory, or the appropriate 4D generalizations.  For instance, it is possible to use the methods of \cite{Coleman:1975pw,Gross:1995bp,Armoni:1997ki,Armoni:2011dw} to compute the $k$-string tension of the 3D theory in the $L\sim N_{c}^{-1}$ large $N_{c}$ limit, as we will discuss elsewhere \cite{ChermanDorigoniForthcoming}.

\section{Acknowledgements}
We are grateful to Adi Armoni, Paulo Bedaque, Mike Buchoff, Francis Bursa, Tom Hammant, Andrei Smilga, Cobi Sonnenschein,  Mithat \"Unsal, Matt Wingate, and Kenny Wong for useful and illuminating discussions.  We are especially grateful to Mithat \"Unsal for detailed comments on a draft of the manuscript.  A.~C. is also particularly grateful to Paulo Bedaque for crucial early guidance and encouragement on this project, and for a stimulating collaboration on related topics.  A.~C. would also like to thank the TQHN group at the University of Maryland for hospitality and support during the early stages of the work.  We thank the Galileo Galilei Institute for providing a fantastic environment for colloboration during the 2011 workshop on ``Large N Gauge Theories'', and are grateful to the INFN  for partial financial support.   D.~D.  is grateful for the support of European Research Council Advanced Grant No. 247252, ``Properties and Applications of the Gauge/Gravity Correspondence", while A.~C. is supported through the DAMTP HEP group STFC grant.

 \bibliography{bosonization,orbifoldingEFT,lightcone}

\end{document}